# The EMPATHIC Project: Building an Expressive, Advanced Virtual Coach to Improve Independent Healthy-Life-Years of the Elderly

*Luisa Brinkschulte[1], Natascha Mariacher[1], Stephan Schlögl[1], María Inés Torres[2], Raquel Justo[2], Javier Mikel Olaso[2], Anna Esposito[3,4], Gennaro Cordasco[3,4], Gérard Chollet[5], Cornelius Glackin[5], Colin Pickard[5], Dijana Petrovska–Delacretaz[6], Mohamed Amine Hmani[6], Ayment Mtibaa[6], Anaïs Fernandez[7], Daria Kyslitska[7], Begoña Fernandez-Ruanova[8], Jofre Tenorio-Laranga[8], Mari Aksnes[9], Maria Stylianou Korsnes[9], Miriam Reiner[10], Fredrik Lindner[11], Olivier Deroo[12], Olga Gordeeva[12]*

This paper outlines the EMPATHIC Research & Innovation project, which aims to re-search, innovate, explore and validate new interaction paradigms and platforms for future generations of Personalized Virtual Coaches to assist elderly people living independently at and around their home. Innovative multimodal face analytics, adaptive spoken dialogue systems, and natural language interfaces are part of what the project investigates and innovates, aiming to help dependent aging persons and their carers. It will uses remote, non-intrusive technologies to extract physiological markers of emotional states and adapt respective coach responses. In doing so, it aims to develop causal models for emotionally believable coach-user interactions, which shall engage elders and thus keep off loneliness, sustain health, enhance quality of life, and simplify access to future telecare services. Through measurable end-user validations performed in Spain, Norway and France (and complementary user evaluations in Italy), the proposed methods and solutions will have to demonstrate usefulness, reliability, flexibility and robustness.

[1] MCI Management Center Innsbruck, Innsbruck, Austria
[2] Universidad del País Vasco UPV/EHU, Bilbao, Spain
[3] Università degli Studi della Campania, Luigi Vinvitelli, Caserta, Italy
[4] International Institute for Advanced Scientific Studies (IIASS), Vietri sul Mare, Italy
[5] Intelligent Voice Ltd., London, UK
[6] CNRS SAMOVAR UMR 5157, Télécom SudParis, Université Paris-Saclay, Paris, France
[7] e-Seniors Association, Paris, France
[8] OSATEK, Bilbao, Spain
[9] Oslo University Hospital, Oslo, Norway
[10] Technion – Israel Institute of Technology, Haifa, Israel
[11] Tunstall Healthcare Ltd., London, UK
[12] Acapela Group S.A., Brussels, Belgium

## Introduction

Much of Europe's elder care is still provided by informal care givers (e.g. family members). Yet, predictions point to less availability for this type of care in the future, for which studies suggest paying more attention to the lifestyle of elderly, helping them remain independent [50]. Psychological and structural obstacles as well as lack of relevant knowledge may however, complicate intentions. Thus, the focus should be on elderly's internal and external difficulties, offering facilities and arrangements, which support active living. According to Kirkwood [28], 75% of longevity may be explained by socio-behavioral conditions the remaining 25% by genetic factors. Based on this assumption, the variability found in environmental factors does to some extent explain variations found in aging; ranging from an active and positive life course to aging with illness and dependence. It is further highlighted that four principles promote active aging. Those four are: (1) *dignity*, which includes privacy, informed consent, autonomy, obtrusiveness and equal access; (2) *autonomy*, whose application requires respect for the self-determination of people and the recognition of heterogeneity in people and their preferences, essentially, in their individual approach to aging; (3) *participation*, which is highlighted by the World Health Organization as one of the pillars of active aging, together with health and safety; and finally (4) *joint responsibility*, which refers to the necessity to involve all generations in the construction of a welfare society supporting all ages. Information and Communication Technologies (ICT) are sought to become an enabler for these 4 principles, helping elderly remain autonomous and independent and consequently, allowing them to stay an active member of their societal community. The EMPATHIC (Empathic, Expressive, Advanced Virtual Coach to Improve Independent Healthy-Life-Years of the Elderly)[13] project aims to significantly contribute to these respective technological advancements.

## Project Overview

Building on initial previous explorations (cf. [31]), the goal of EMPATHIC is to research, innovate and validate new interaction paradigms, laying the foundation for future generations of Personalized Virtual Coaches that help elderly people live independently. The envisioned wellness coaching advice seeks to promote healthy habits and behavior, by challenging users to transform their personal goals and needs into actions. Furthermore, the EMPATHIC Virtual Coach (EMPATHIC-VC) aims to engage its healthy senior users so that they actively reduce their risks with respect to potential chronic diseases, that they maintain a healthy diet, have adequate physical activity, as well as encourage social engagement, thus contributing to their ability to maintain a satisfying and independent lifestyle. The EMPATHIC-VC shall motivate users through a number of project-defined metrics whose ambition is to create a personal, friendly and familiar environment, avoiding the threatening effects of unfamiliar new gadgets or an excessive focus on medical supervision. As such, the project seeks to set its focus beyond the basic medical and physical needs of elderly, and rather create a link between their body and emotional well-being. The EMPATHIC-VC shall be capable of perceiving emotional and social states of people, learn and understand senior users' context dependent expectations and requirements, as well as their personal history, and thus respond adaptively to their needs. To achieve these goals, EMPATHIC works on innovative, multimodal face and speech analytics, adaptive spoken dialogue systems, intelligent computational models and natural human-computer interfaces, supporting the development of an emotionally-expressive virtual coach, designed to help aging users and their carers. Building upon neuroscience research, the project will use unobtrusive remote technologies so as to extract physiological markers of emotional states in real-time. That is, the EMPATHIC-VC shall monitor facial cues and speech style, which underpin users' basic neural functions, and from this formulate online adaptive responses, facilitating interaction through mimicking and in turn promoting empathy in the user. To develop the proposed EMPATHIC-VC, a number of scientific and technological challenges need to be addressed:

1. The virtual coach has to listen to users and correctly understand the information communicated by their face, speech and language, through non-intrusive technologies;
2. It has to be capable of deriving the personal needs of users from a history of user-coach interactions, as well as identify users' emotional status and concerns;
3. It has to be able to implement high level healthy-aging and well-being plans, and construct an effective motivational model, implemented by multimodal dialogues specifically tailored towards individual users.

---

[13] https://www.cordis.europa.eu/project/rcn/212371_en.html

In order to maximize the acceptability, use and impact of the EMPATHIC-VC, it will be designed so that it runs primarily on tablets and smartphones, but also PCs, laptops, and even TV screens may be supported if such is highlighted by user preferences. To ensure user-driven design, EMPATHIC will involve senior users throughout all phases of the project, considering their input with respect to functional and non-functional product requirements. All the developed technologies will respect data privacy in terms of secure storage, management and transmission according to EU and national regulations. They will be demonstrated and validated in a series of clearly-defined realistic use cases representing different coaching goals and user profiles. The respective test sample will consist of 200-250 senior users recruited in three European countries (i.e. Spain, Norway and France), representing different culture and lifestyle environments.

## Project Objectives

The EMPAHIC research focuses on **six objectives**. The **first objective** deals with the design of the EMPATHIC-VC, which shall engage healthy-senior users and reach preset benefits, measured through project-defined metrics, to enhance well-being. We aim to achieve this by using the EMPATHIC-VC to increase awareness of one's personal physical status, to improve diet and nutritional habits, and to actively promote physical exercises and social activities. **Objective two** focuses on the continuous involvement of end-users, so as to optimally align the technology with the users' personalized needs and requirements, all of which shall be derived by the coach. The **third objective** aims to supply the coach with incorporated, non-intrusive, privacy-preserving, empathic, and expressive interaction technologies. **Objective four** centers around the validation of the coach's efficiency and effectiveness across three distinct European societies (i.e. Norway, Spain and France), evaluated by 200 to 250 end users.

**Objective five** evaluates and validates the effectiveness of the EMPATHIC-VC designs against users' very personalized acceptance and affordance criteria (such as the ability to adapt to users' underlying mood). Finally, **objective six** concerns the EMPATHIC-VS's industry acceptance and open-source access, identifying appropriate evaluation criteria to improve the 'specification-capture-design-implementation' software engineering process of implementing socially-centered ICT products. These six objectives will be tackled by the following research and development actions:

- First, we will provide automatic, personalized advice guidance through advanced ICT. That is, we will investigate the identification and assessment of main cues related to physical, cognitive, mental and social well-being.
- Second, we will define personalized, psychologically motivated and acceptable coach plans and strategies.
- Moreover, the project will research the translation of professional coaching behavior into actions to be performed by the intelligent EMPATHIC-VC.
- Also, research will investigate non-intrusive technologies to detect markers for individuals' emotional and health status. For this, emotional information from eyes, face, speech and language will be used to deliver a hypothesis and consequently assist the decisions to be taken by the EMPATHIC-VC.
- A focus is also investigations into sudden shifts in users' emotions. Here, planned efforts span around the implementation of discrete health-coach goals and actions through an intelligent computational system, underlying a spoken dialogue system adapted to users' intentions, emotions and context. Consequently, research focuses on data-driven modelling of users and tasks, machine learning for understanding users, on learning policies and questionnaires to deal with coach goals, on statistical approaches for dialogue management driven by user and coach goals, and on online learning for adaptation.
- Another concrete work action is to provide the EMPATHIC-VC with natural, empathic, personalized and expressive communication capabilities, which supports emotional bonds that result in engaged and effective relationships.

Three sets of KPIs will be used to route the fulfillment of these objectives. First, **usefulness and effectiveness** in quality of life for the coached elderly will be consistently analyzed. Second, **task success** both with respect to the VC's system decisions and the Spoken Dialogue System's understanding, will be measured. Finally, **user acceptance and usability** as well as **cost-effectiveness** will be evaluated.

## Relevant Related Work

Research areas where the EMPATHIC project aims to offer advances include virtual health and wellness coaching, understanding users' speech, extracting emotions from spoken language, assessing user preferences when interacting with machines, making decisions in Spoken Dialogue Systems, defining natural, expressive verbal system interactions, and adapting talking-face agents to user preferences. The following sections provide an overview of these areas.

## Virtual Health and Wellness Coaching

Due to a predicted future lack of informal caregivers, attention needs to be put on counteracting the consequent challenges for elderly. The idea to offer virtual health and wellness coaching draws on theories of human development, social psychology, organizational leadership, and adult learning theory, where the term "Virtual Coach" does refer to a coaching program or device aiming to guide users through tasks for the purpose of prompting positive behavior or assisting them with learning new skills. While from a pure utilitarian perspective the coach's challenge is to support users in improving their lifestyle, delivering relevant messages empathically, rather than simply stating facts, triggers emotions associated with reaching a goal and thus help promote a trustworthy relationship between a user and the system [18]. In contrast to, for example, a medication reminder, virtual coaches monitor how users perform activities, provide situational awareness and provide feedback and encouragement that matches users' cognitive states and circumstances [49].

## Understanding Users' Speech

One recurring problem of Automatic Speech Recognition (ASR) concerns the dependence on transcribed speech data where phonemes, words and other speech units are accurately timed. Additionally, the production of an ASR model for a particular language requires the time-consuming development of language models. Consequently, the focus of developers has been on the development of ASR systems, which neither require explicit language models nor the accurate timing of speech units. One bottom-up approach to this is Automatic Language Independent Speech Processing (ALISP) [14], where spectral features corresponding to speech units are extracted automatically from acoustic data. A more recent top-down approach relies on the availability of large amounts of speech data with text transcripts such as movie scripts and subtitles. Whilst these transcripts do not contain timing of speech units at the word or phoneme level, they can be aligned with spectrogram slices using connectionist temporal classification (CTC) [21]. The application of Recurrent Neural Networks (RNN) to language is facilitated by the recent appearance of vector representations of words, such as Word2Vec [38] and GloVe [43], trained using Skipgram auto-encoders and their inverse Continuous Bag of Words (CBOW) approaches. Traditional GMM-HMM based ASR systems are essentially models of sequences of acoustic states, with state probabilities optimized using statistical language models. However, for large models accurate estimation of probabilities of sequences of states becomes increasingly difficult. Hence, language models based on RNNs, which began with pioneering work by Bengio [5], are now becoming more mainstream [37]. This is crucial for many temporal applications, but particularly for speech and natural language processing where long-term dependencies in characters or words need to be exploited for natural language parsing of sentence structure, context and meaning. Innovations of RNN network architectures such as the emergence of sequence-to-sequence models [13], which have encoder and decoder RNN components connected together using a so-called neural attention mechanism, were developed for machine translation applications but are now being used for a range of sequence-based tasks [54]. These deep learning approaches are changing the landscape of understanding a user's speech.

## Extracting Emotions from Spoken Language

The interaction between human beings and computers shall become more natural once computers are able to perceive and respond to humans' non-verbal communication cues, such as those triggered by emotions. Emotion recognition from speech, face images and video have a long history. In recent years, researchers began to exploit diverse signals to analyze human emotions. Busso [8], for example, showed that emotion recognition improves considerably using facial expressions, speech and multimodal Information. Soleymani [50] on the other hand, presented a user-independent emotion recognition method, which aims at recovering affective tags for videos using electroencephalogram (EEG), pupillary response and gaze distance. In general, eye tracking has become an optimal technology for studying cognition, and human emotion correlates. Thus, multimodal emotional databases begin to be available [35], and new algorithmic methods, which exploit such multimodal emotional data with deep neural networks [26] for emotion recognition, start to appear.

## Assessing User Preferences when Interacting with Machines

Europe counts an ever-increasing number of senior citizens suffering from cognitive decline, different types of mental disorder, anxiety or depression[14]. Additionally, it is estimated that 47.5 million people worldwide are living with dementia – a number that is predicted to further increase in upcoming next years[15]. This places considerable burdens on national health care institutions in terms of medical, social, and care costs [34]. Information and Communication Technology systems, in the form of embodied conversational agents, have been proposed to reduce these burdens because they can provide automated on demand health assistance, thereby reducing the above-mentioned costs (cf.

---

[14] www.parkinsons.org.uk/content/about-parkinsons, www.who.int/mediacentre/factsheets/fs381/en/, Munoz, 2010
[15] http://www.who.int/mediacentre/factsheets/fs381/en/

Hawley et al. for communication disorders [23], Parker for ageing [42], and Prescott for companionship [44]). Thus, currently there is a huge demand for complex autonomous systems able to assist people with their needs, ranging from long term support of health states (including caring for elders with motor-skill restrictions) to communicative disorders. Provisions of support have been made through both the monitoring and detection of changes in the physical, and/or cognitive, and/or social daily functional activities, as well as in offering therapeutic interventions [46]. Until now, however, there have been relatively few efforts assessing human interactional exchanges in context in order to develop systems that are able to establish a trusting relationship with users [10], take appropriate and autonomous actions to provide long-term and meaningful help [17, 9], and rise emphatic feelings so as to build up therapeutic relationships with users.

## Making Decisions in Spoken Dialogue Systems

Spoken dialogue systems have been developed in various domains and dealing with different goals. Typically, they provide users with information on topics such as flight schedules [3], restaurant recommendations [52], tourist attractions [27] or customer service [48]. Particularly relevant to EMPATHIC are spoken dialogue systems supporting tele-medical and personal assistant applications for older people [41]. The basic tasks of a dialog system is to interpret the user's input (Automatic Speech Recognition) in a given context (Natural Language Understanding), to decide what action the system should take (Dialogue Management), and consequently produce an appropriate response (Natural Language Generation). The input speech is transformed into text using Automatic Speech Recognition (ASR), and the output forwarded to a Text-to-Speech Synthesizer. The Natural Language Understanding (NLU) component identifies the intent of the speaker and the semantic content. This classification can be performed using either a data-driven statistical [33] or a knowledge-based approach, (e.g. through handcrafted grammars). The major challenge is the identification of a semantic inference based on often ambiguous natural language input. Thus, the Dialogue Manager (DM) decides upon the action a system should perform at any given state in the dialogue, using the string of semantic units provided by the NLU as well as the (learned) history of the dialogue. The applied supervised learning is focused on making predictions about inputs [7, 22, 4]. When building systems for virtual coaching, there is, however, a need to build personalized models for each user. One possible approach is to use data collected from all users and consequently build a general user model. Another approach focuses on building individual models per user, yet such often induces noise, due to insufficient data. A framework used to balance these two extremes is the so-called multi-task framework. Here, all data is used to build a number of models based on either clustering [3, 15, 25], sharing parameters [36, 11, 20], or features [30, 1].

## Defining Natural, Expressive Verbal System Interaction

Over the last 50 years, Text-to-Speech (TTS) technology has evolved, starting from producing rather robotic voices [29] to more humanoid voices in the 90s [12], until at the beginning of the 21st century eventually more sophisticated and natural voice technology started to appear [24]. The next breakthrough in this domain should thus go beyond pure TTS optimization and rather start including human characteristics in dialog, such as for example relevant social cues in discussions. These types of advancements, however, require the technology to have the ability to render emotions, to use different prosody models, or even to understand the meaning of text and thus adapt the speech accordingly.

## Adapting Talking-Face Agents to User Preferences

Virtual representations of humans in computer mediated interactions are usually categorized as avatars or agents. Avatars are controlled by humans, whereas agents are controlled by computer algorithms. Hence, interaction with an avatar qualifies as computer-mediated communication [39], whereas interaction with an agent qualifies as human-computer interaction. Several studies have emphasized the importance of building agents and avatars able to create a relationship with users while expressing emotions [6]. Agents showing appropriate emotions lead to higher perceived believability and trust. Agents have also been endowed with conversational skills able to display communicative gestures, and they can adapt their facial expressions depending on their social relationship with interlocutors [40]. Yet, these models work only in specifically controlled contexts.

## Proposed Methodology

The EMPATHIC technology development aims at being end-user driven from the very beginning. 200-250 seniors will be recruited in three European countries, i.e. Spain, Norway, and France, representing different culture and lifestyle environments. Developments will run along the following key stages:

- User-driven requirements analysis and selection of use cases
- Definition of architecture
- Privacy-driven collection of initially required language resources
- Mid-period prototypical proof-of-concept
- Technology Integration
- Privacy-driven end-user validation in 3 countries
- Stakeholder inclusion, dissemination and exploitation planning, including operative recommendations

## EMPATHIC-VC Outline

The EMPATHIC-VC aims to combine different functional aspects. The agent should represent an artificial entity with machine/deep learning capabilities, and online access to a database holding social, medical and administrative history information on users, which will be used to provide intelligent coaching. To build upon this, the coach should further be equipped with emotions that humans can read. Humans read emotions automatically, independently of education or intelligence. It is not sufficient that the EMPATHIC-VC only extracts and understands emotions. Rather it needs to have the intelligence to extract the event related facial-voice-eyes-posture states and generate emotions in a way that mirrors humans in expressing their emotions. To achieve such capabilities, it is planned to use an existing database of emotional faces, which go beyond the seven Ekman expressions [19], and couple this with voice frequencies and pitch to validate humans' emotional perception. While a set of personalized variables and parameters will represent the initial configuration of the EMPATHIC-VC, their values and influence will be consistently modified through learning algorithms driven by real interactions.

## EMPATHIC-VC Architecture

The EMPATHIC-VC will be designed as a client-server architecture, where the client is running in the form of a (web-based) application on tablets, smartphones, laptops or PCs. The device's own camera and microphone will be used to sense a user, the display to present a talking face agent. Figure 1 shows the proposed EMPATHIC-VC architecture.

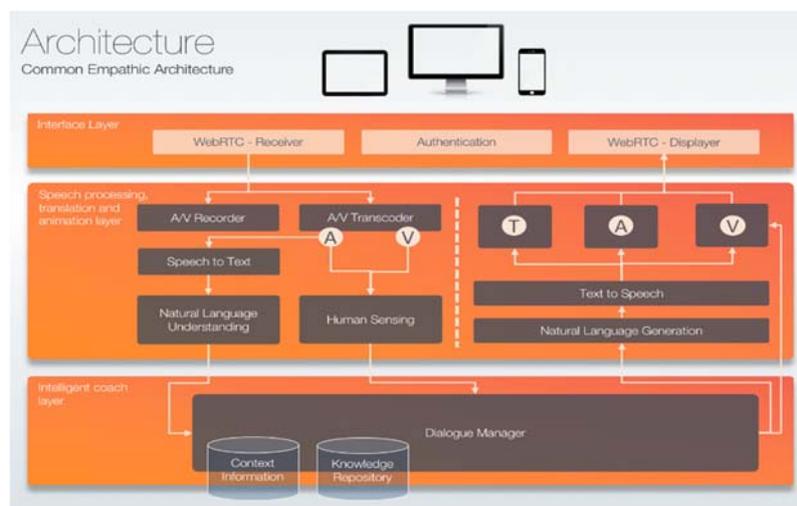

**Figure 1: Proposed EMPATHIC-VC Architecture**

The web frontend will relay audio-visual information between the client and server using WebRTC. The authentication and authorization service recognizes the user and provides application access. The expressive translation service is responsible for translating natural language text and emotional cues into an inbound semantic entity stream, and outbound semantic entity streams into text and emotions. The dialogue manager is responsible for constructing and dispatching queries to the intelligent coach system based on the current inbound semantic entity stream, i.e. the information obtained from the context provider service. The dialogue manager also compiles the information received from the intelligent coach into an outbound semantic entity stream, which is dispatched back to the translation service. The dialogue history service stores and supplies relevant information collected from previous dialogues with a user. The intelligent coach service is responsible for responding to queries from the dialogue manager and uses the user's knowledge bank and general knowledge bank services. The user's specific knowledge bank is a data store containing specific information about the current user. The general knowledge bank stores medical data, workflows and actions. The semantic entity stream is a stream of semantic information represented in a format that facilitates computational operations and queries. The context provider service persists and supplies contextual information about an ongoing dialogue. When the dialogue has ended this service extracts the relevant information and supplies contextual information to the dialogue history service, after which all contextual information is deleted.

As seen in the Figure 1 the application components are grouped into three distinct layers. The first layer is the interface layer, which identifies and authorizes users of the system, and selects suitable communication endpoints based on the location of the user's device, to minimize latency between the client and the server application. The next layer is the speech processing, translation and animation layer, which receives and submits audio and video using two-way translation into semantic entity streams to communicate with the intelligent coach layer. Finally, the intelligent coach layer uses the available knowledge and data to create dialogue strategies and actions.

Figure 2 illustrates the data flow in the EMPATHIC network presenting main hardware and software components. It shows where hardware servers should be located in relation to their software components in order to reflect the volume of computation or hardware specialism the servers should have.

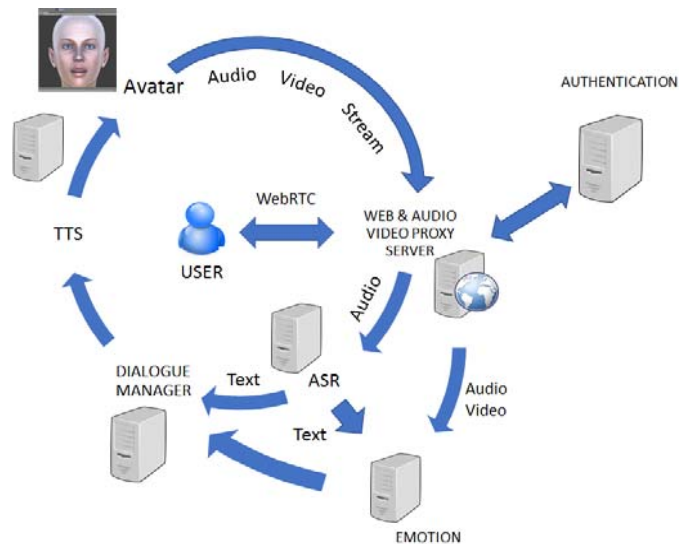

**Figure 2: Data flow in the EMPATHIC network**

In order to simulate important system functions, which require ongoing development (e.g. speech recognition and language understanding), a Wizard of Oz (WOZ) component will be developed. The WOZ component will be used to explore user responses and the consequent handling of the dialogue, to test different dialogue strategies and to collect language and video resources needed to train technology components. Furthermore, the WOZ component will allow for collecting additional feedback and preference data related to the appearance of the agent, helping its ongoing augmentation.

## EMPATHIC-VC Evaluations

The EMPATHIC-VC will be tested by representative, realistic use cases for different user profiles, in three different European countries, i.e. Spain, France and Norway (complemented by additional evaluations in Italy). A sample of 200-250 users will allow for the evaluation of various scenarios with respect to the earlier presented project KPIs. Two specific scenarios, including a communication and a daily care scenario, will be explored. The first scenario counts as a free scenario where the user may ask about the weather, restaurants, movies and other miscellaneous information. The second scenario is agent-driven and inquires on user feelings, health states, and dietaries. In cases where the technological progress of the EMPATHIC-VC is not ready to satisfy interactional requests, the WOZ component will take over. After each interaction, post-test interviews will be conducted, focusing on users' subjective judgements about the quality of their interaction with the system, the interaction's easiness/ difficulty, the quality of the lived experience, and the expected benefits when requiring the system's services in each scenario.

The outputs of these evaluation will include (1) a database of audio-video recordings of user-system interactional exchanges; (2) a lexicon of quantitative and qualitative usability metrics; (3) parameters describing the appearance efficacy and trustworthiness of the system; (4) a set of qualitative and quantitative parameters describing users' needs; (5) expectations and requirements related to three different user-systems interaction modes; and (6) a mathematical model of multimodal interactional factors serving as tool for identifying explicit and implicit interactional preference features for establishing empathic therapeutic relationships between the EMPATHIC-VC and its users.

All interactions will be video-recorded so as to analyze user behavior, emotional, vocal, facial, and gestural expressions, as well as, the flow of the dialogues. Furthermore, participants will be asked to assess their individual impression of the system as well as their perceived ease of use and effectiveness, their expectations and their overall acceptance of the system. The recorded interactions will be assessed with respect to their degree of empathy and trust through the Interpersonal Reactivity Index (IRI) [16] and the Barrett-Lennard Relationship Inventory (BLRI) [2]. The Rotter scale [47], Rempel and Holmes scale [45] will be used to measure participants' tendency to trust others and their trust level towards a specific conversant. This type of analysis and the mathematical modelling of the interactional parameters, through the combined processing and the cross-linguistic and cross-cultural comparisons of the three field-trial languages, shall define degrees of preference, acceptance and trustworthiness for the developed system. The result will be a mathematical model of multimodal interactional factors serving as a tool for identifying elder's explicit and implicit preferences for interface modalities. First, the model's validity will be assessed for each single factor (e.g. for the interaction levels) and successively extended to define different integrated parameter sets. The best fit will then be used as elders' system preference identification tool. Next, a Confirmatory Factor Analysis (CFA) [53] will test whether the correlational structure of the identified factors is consistent with the proposed model. Furthermore, a two-phase validation study will collect caregiver's judgment with respect to the system's feasibility, effectiveness and usability.

## Conclusion

The paper outlined the goals and methods proposed by the EMPATHIC Research & Innovation project. The envisioned EMPATHIC Virtual Coach for elderly aims to advance research related to emotion extraction and speech recognition as well as to dialog management, human-computer interaction and virtual health and wellness coaching. The project is strictly committed to an end-user driven, culture sensitive development approach, which focuses on realistic use case scenarios to be tested and evaluated with 200-250 seniors in three European countries (i.e. Spain, France and Norway). In doing so, EMPATHIC seeks to lay the foundation for future generations of Personalized Virtual Coaches to assist elderly people living independently at and around their home.

## Acknowledgment

The research presented in this paper is conducted as part of the project EMPATHIC that has received funding from the European Union's Horizon 2020 research and innovation programme under grant agreement No 769872.

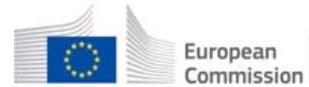